\documentclass[prl,twocolumn,showpacs,preprintnumbers,amsmath,amssymb,ti
ghtenlines]{revtex4}

\usepackage{graphicx}
\usepackage{dcolumn}
\usepackage{bm}

\begin{document}

\preprint{\today}
\title{Limit to non-destructive optical detection of atoms}
\author{J.J. Hope and J.D. Close}
\affiliation{Australian Centre for Quantum Atom Optics,
Department of Physics,
Australian National University,
ACT 0200, Australia}

\email{joseph.hope@anu.edu.au}

\begin{abstract}
All optical techniques used to probe the properties of Bose-Einstein
condensates have been based on dispersion and absorption that can be
described by a two-level atom.  Both phenomena lead to
spontaneous emission that is destructive.  Recently,  both were shown to
lead to the same limit to the signal to noise ratio  for a  given destruction.
We generalise this result to show that no single-pass optical technique using
classical light, based
on any number of lasers or coherences between any number of levels, can
exceed the limit imposed by the two-level atom.  This puts significant
restrictions on potential non-destructive measurement schemes.
\end{abstract}

\maketitle

\textit{Introduction}.- The advent of modern cooling techniques has led
to the creation of ultra-cold atomic samples in which the recoil of a
single photon has a significant effect on the motional state of the
system.  Laser cooling, and more recently evaporative cooling, have
allowed the creation of a Bose-Einstein condensate (BEC) of weakly
interacting gases, in which a large number of atoms enter the ground
state of the system, forming a large, coherent matter wave
\cite{BECreview}.  Observation and control of the motional states of
these atoms requires a detection method that does not involve
spontaneous photon recoil.

Previously all ground state BECs have been
detected via optical methods, with photon absorption providing a
simple, though clearly destructive, measure of the atomic density and
the phase shift of a laser beam providing a less destructive measure
under some circumstances \cite{Ketterle1999}.  Both methods are based
on physics that can be described by the two-level atom.  It was
recently shown in the limit of optically thin samples, that absorption
and phase shift measurements have equal sensitivity for a given level
of destruction, and that the signal to noise ratio (SNR) in this limit
is a function of destruction (spontaneous emission rate) and bandwidth
only \cite{Lye03}. In this way, there is a hard limit on the SNR achievable 
from any single-pass technique based on the two level
atom and classical light beams.

It is known that in three-level systems
in the presence of a strong second laser, it is possible for a weak
probe beam to experience a non-zero phase shift without any absorption,
suggesting that manipulation of coherences in a three-level system
might provide a less destructive detection method
\cite{Arimondo,Scully}. In this Letter we show this is not the case
and, further, that no single-pass optical detection method can beat the
limit imposed by the two-level atom. The only solution is to multi-pass
the beam with a cavity or to use non-classical light.

The phase shift of a far-detuned laser beam passing through a
gas  of two-level atoms is inversely proportional to the detuning. The
excited
state population and accompanying spontaneous emission is inversely
proportional to the square of the  detuning. As a consequence it might be
thought that detuning further from resonance will
provide improved sensitivity for a given excited state population.
Careful analysis shows, however, that the SNR for the shot noise
limited measurement of a phase shift is also proportional to the
electric field amplitude of the beam. Consequently, the SNR is
proportional to the square root of the excited state population with no  
free parameters that can improve the performance of the system.  In the
limit of an optically thin sample, the prefactors are in fact identical
to that of an absorption measurement.  There are three ways  of beating
this limit with two level atoms, but they have either limited application
or impose significant
technical difficulties for moderate gains using current technology.
Measuring the resonance fluorescence can
provide a fixed improvement in the ultra-thin limit where both the SNR
and the excitation go to zero.  Using resonant interferometry provides
a factor of the square root of the finesse of the system \cite{Lye03}.
Using squeezed light improves the SNR by a
factor of the squeezing.  These technically challenging methods provide
techniques by which the SNR can be enhanced \cite{Hinds}, but would
not be worth pursuing unless the limit imposed by the two-level atom
were fundamental to all  single-pass optical detection schemes using
classical light.

The properties of dark states alluded to earlier (absorption-free  phase
shift  of a probe) suggest that a three-level system  could provide  a
less destructive detection method.  The correct measure of  destructiveness
is not, however, the total absorption of the probe beam, but the
 total spontaneous emission rate due to all laser beams or, succinctly, the
excited state population. For the three-level lambda system,
there is a dark state that has exactly no excited state population, but
it also gives no phase shift to the laser beams. To determine whether the
two-level limit can be beaten by such a scheme, we must
calculate both the phase shift on the probe and the excited state populations for all
states.  The
relationship between the phase shift on any laser beam in a multiple
laser, multiple level system and the excited state population can be
determined in various limits, but in order to determine whether there
is any non-destructive scheme this must be calculated without resorting
to linearisation of the susceptibility or other perturbative methods.


\textit{Non-perturbative phase shift in multilevel systems.}-- The phase
shift on a laser beam can be determined using a semiclassical
analysis and the wave equation, but this has a tendency to become
extremely complicated in the presence of non-linear susceptibilities
and multiple transitions.  The fully quantum method described in this
Letter is a surprisingly efficient computation of the phase shift where
the non-linear dynamics is entirely included in the calculation of the
dressed state energies of the atoms.  The phase
shift on a laser beam interacting with a complicated system is easiest
to calculate by identifying its origin in the level shifts of the
dressed states of the system.  Let us describe the initial atomic state
as $|1\rangle$, which will be connected to a series of other atomic
states $|j\rangle$ by optical fields that are each in a coherent state
$|\alpha_k\rangle$.

\begin{figure}
\includegraphics[width=7.5cm]{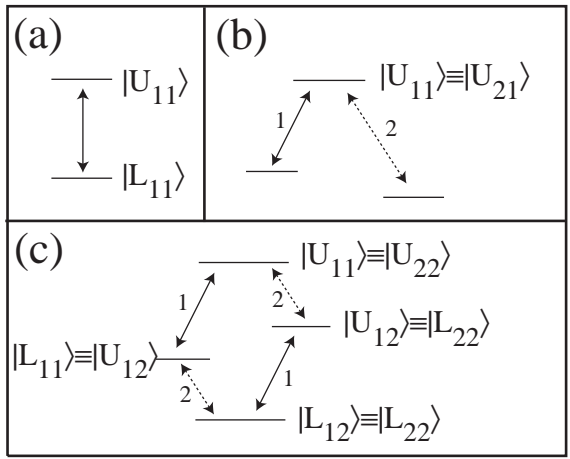}
\caption{Examples of lasers and atomic energy level schemes that are
allowable in our analysis.  (a) shows the two-level atom.  This can be
connected only by a single laser, or else it would be possible to
return to the ground state without returning to the original momentum
state.  (b) shows a Raman transition with two laser fields.  Each laser
still causes only one transition, but the excited state for each of
those transitions is identical, so in our notation $|U_{1
1}\rangle\equiv|U_{2 1}\rangle$.  (c) shows a more exotic system where
each laser causes multiple transitions but the atoms will still return
to the starting state without momentum diffusion.  In this example, all
four states are multiply defined in the $|U_{j l}\rangle, |L_{j
l}\rangle$ notation.}
\label{fig:levelexamples}
\end{figure}

While we do not wish to make assumptions as to the nature of the system
we are investigating, our analysis is simplified  in that we
are looking for a non-destructive detection scheme, and we can
immediately rule out combinations of lasers and level systems that lead
to a net change in the atomic electronic state or momentum after the
interaction with the lasers.  The combined system of atoms and lasers
must therefore reduce to a series of manifolds in which each atomic
level is associated with a definite number of photons in each optical
mode, or else the atoms will experience momentum diffusion during their
interaction with the lasers.  For a two-level system, for example,
only one laser can couple the states $|1\rangle$ and $|2\rangle$.
During the interaction, the total quantum state will reduce to a set of
non-interacting manifolds of pairs of states $\{|1,n\rangle,
|2,n-1\rangle\}$, where $|j,n\rangle$ is the state with the atom in
electronic state $j$ with $n$ photons in the laser beam
\cite{Tannoudji}.  A three-level system in a lambda configuration where
lower energy states $|1\rangle$ and $|3\rangle$ are each coupled to the
excited state $|2\rangle$ by a separate laser mode is another example.
In this case the independent manifolds are $\{|1,n,m\rangle,
|2,n-1,m\rangle, |3,n-1,m+1\rangle\}$.

In general, the closed manifolds can be indexed by the atomic state
alone, although the full quantum state will include the number of
photons in each optical mode, $\{|j,n_1+b_{1 j},n_2+b_{2 j}, \cdots,
n_M+b_{M k}\rangle\}$, where $b_{i j}$ are the elements of an
integer-valued, constant matrix.

The interaction picture Hamiltonian for such a system with $N$ atomic
levels and $M$ different lasers can be written, after the rotating wave
approximation, as
\begin{eqnarray}
H & = & \sum_{n=1}^N{\hbar \Delta_n |n\rangle\langle
n|}+\sum_{j=1}^N\sum_{l=1}^{m_j}(g_{j,l} \hat{a}_j^\dag |L_{j
l}\rangle\langle U_{j l}| + \mbox{adj.})
\nonumber
\end{eqnarray}
where $\hbar \Delta_n$ is the energy of atomic level $n$, $\hat{a}_j$
is the annihilation operator for the optical mode of the
$j^{\mbox{th}}$ laser, $m_j$ is the number of transitions caused by the
$j^{\mbox{th}}$ laser, $|L_{j l}\rangle$ and $|U_{j l}\rangle$ are the
lower and upper atomic energy levels respectively of the appropriate
laser transition, and $g_{j,l}$ is its dipole coupling strength.  Examples of 
systems with different numbers of atomic levels and lasers are shown in 
Fig.(\ref{fig:levelexamples}).

The requirement that the atoms be left in their starting electronic
state can be solved either by requiring that the lasers perform a
multi-level equivalent of a $2\pi$ pulse, or by assuming that the atoms
adiabatically follow the dressed states.  For a pulsed system, the phase
shift vanishes.  For the second case, we have found an efficient
method of calculating the phase shift for a system with an arbitrary
number of atomic levels and incident lasers.

If the atoms adiabatically follow the dressed state and the photon
number of each laser is conserved, the final state can differ at
most by a phase factor that is the product of the eigenvalue of the
relevant dressed state,
$\langle H \rangle(n_1,\cdots,n_M)$, and the time $\tau$ of the
interaction.  If this total phase factor is linear in each photon
number, $|j,n_1+b_{1 j},n_2+b_{2 j}, \cdots, n_M+b_{M k}\rangle
\rightarrow |j,n_1+b_{1 j},n_2+b_{2 j}, \cdots, n_M+b_{M k}\rangle
e^{-i (\sum_j \Delta\phi_j n_j) \tau}$,  the total state is
trivially separable into the outer product of coherent states each with
an associated phase shift $\Delta\phi_j$.  This is true for the
two-level atom interacting with a single laser, but will not be true in
general.  The effect of an atomic medium on the light
fields can be more complicated than a simple phase shift on each beam,
even in the absence of absorption or spontaneous emission.  Lasing
modes have relatively well defined photon number and
linearisation of the dressed state eigenvalue around those photon
numbers  yields  a reasonable approximation of the final state.  To the
extent that
the laser fields have
something as simple as a phase shift, it is given by:
\begin{eqnarray}
\left. \Delta \phi_j \right|_{\mbox{atom}} & = & -\frac{l}{\hbar c}
\left.\frac{\partial \langle H \rangle}{\partial
n_j}\right|_{(n_1,\cdots,n_M)=(\bar{n}_1,\cdots,\bar{n}_M)}
\nonumber
\end{eqnarray}
where $\bar{n}_j$ is the average photon number in the $j^{\mbox{th}}$
laser, $l$ is the length of the interacting region and $c$ is the speed
of light.

Within each manifold, the Hamiltonian is an $N \times N$ matrix:
\begin{eqnarray}
H & = & \sum_{n=1}^N{\hbar \Delta_n |n\rangle\langle
n|}+\frac{\hbar}{2} \sum_{j,l=1}^{N,m_j}(\Omega_{j,l} \;|L_{j
l}\rangle\langle U_{j l}| + \mbox{adj.})
\nonumber
\end{eqnarray}
where $\hbar \Omega_{j,l}=2 g_{j,l} \sqrt{n_j+1+b_{j,L_{j l}}}$ is a
shorthand for the resultant coupling term that can be identified as the
resonant Rabi frequency of the transition, making a simple connection
with the semiclassical picture.  The phase shift per
atom on each laser beam is:
\begin{eqnarray}
\left. \Delta \phi_j \right|_{\mbox{atom}}& = & -\frac{l}{\hbar c}
\sum_{l=1}^{m_j}\frac{\partial \langle H \rangle}{\partial
\Omega_{j,l}} \frac{\partial \Omega_{j,l}}{\partial n_j}
\nonumber
\end{eqnarray}
Multiplying by the total number of atoms in the quantisation
volume, we find the total phase shift on the laser:
\begin{eqnarray}
\left. \Delta \phi_j \right|_{\mbox{total}}&=&-\sum_{l=1}^{m_j}
\frac{\tilde{n} \;\sigma_{j,l}\; \gamma_{j,l}}{2 \;\hbar
\;\Omega_{j,l}} \frac{\partial \langle H \rangle}{\partial \Omega_{j,l}}
\nonumber
\end{eqnarray}
where $\tilde{n}$ is the column density of the atoms, $\sigma_{j,l}=6
\pi/k_j^2$  is the single atom cross-section, and $\gamma_{j,l}$ is the
spontaneous emission rate of the excited state $|U_{j l}\rangle$.  The
derivative of the dressed state eigenvalue can be found from first
order perturbation theory using the Hellman-Feynman theorem
\cite{Merzbacher}.  As the Hamiltonian is linear with respect to $\Omega_{j,l}$:
\begin{eqnarray}
\frac{\partial \langle H \rangle}{\partial \Omega_{j,l}}&=&\langle
\Psi_{\bar{n}_1,\cdots,\bar{n}_M}| \frac{\hbar(|L_{j l}\rangle\langle
U_{j l}|+
|U_{j l}\rangle\langle L_{j l}|)}{2}
|\Psi_{\bar{n}_1,\cdots,\bar{n}_M}\rangle
\nonumber
\end{eqnarray}
and we can write the total phase shift in terms of the real parts
of the off-diagonal density matrix elements:
\begin{eqnarray}
\left. \Delta \phi_j \right|_{\mbox{total}}&=&-\sum_{l=1}^{m_j}
\frac{\tilde{n} \;\sigma_{j,l}\; \gamma_{j,l}}{2 \;\hbar
\;\Omega_{j,l}} \Re\{\rho_{L_{j l} U_{j l}}\}.
\nonumber
\end{eqnarray}

A measurement of this phase is therefore a measurement of column
density.  For a purely shot-noise limited measurement without using
squeezing or a resonant cavity, the maximum achievable SNR is limited
by the temporal and spatial bandwidth, detector efficiency and the
strength of the electric field in the interferometer:
\begin{eqnarray}
\mbox{SNR}_j&=&\sqrt{\frac{\eta \;P_{j}}{B \;\hbar \;\omega_{j}}}
\left|\Delta\phi_{j}\right|
\nonumber
\end{eqnarray}
where $P_j$ is the power in the $j^{\mbox{th}}$ laser mode, $\eta$ is
the quantum efficiency of the detector and $B$ is the temporal
bandwidth of the measurement.  The square root of the power and the
inverse Rabi frequency in the phase shift have opposite dependence on
the electric field, and we obtain a SNR that depends only on fixed
atomic parameters and the off-diagonal elements of the density matrix
of the relevant dressed state.
\begin{eqnarray}
\mbox{SNR}_j&=&\sum_{l=1}^{m_j} \frac{\tilde{n}}{2}  \sqrt{\frac{\eta
\;A\;\sigma_{j,l}\; \gamma_{j,l}}{B}} \Re\{\rho_{L_{j l} U_{j l}}\}
\nonumber
\end{eqnarray}
where $A$ is the area of the atoms that was sampled - essentially the
spatial bandwidth of the measurement.

Using basic properties of the density matrix we can immediately write
this as an inequality:
\begin{eqnarray}
\mbox{SNR}_j&\le&\sum_{l=1}^{m_j} \frac{\tilde{n}}{2}  \sqrt{\frac{\eta
\;A\;\sigma_{j,l}\; \Gamma_{j,l}}{B}} \label{eq:limit}
\end{eqnarray}
where $\Gamma_{j,l}=\gamma_{j,l} \;\rho_{U_{j l} U_{j l}}$ is the
spontaneous emission rate per atom due to the population $U_{j l}$ in
the upper state of the $l^{\mbox{th}}$ transition due to that laser.
This new result shows that there is a fundamental limit to the
sensitivity for any coherent optical detection method \textit{for a
given level of disruption of that state}.

\textit{Relationship to other theorems}.-  It is important to put this
work in perspective with other general theorems in optics. The
Kramers-Kronig (KK) relations relate the imaginary part of the
refractive index of a gas at a particular frequency to the integral of
its real part over all frequencies, and vice versa \cite{Jackson}.  In
contrast, equation (\ref{eq:limit}) addresses the achievable SNR in a
quantum noise-limited measurement for fixed total absorption at a
particular frequency.  Although superficially related, the two are
quite different. The KK relations relate purely classical aspects of the
fields in a way that is not applicable to the question of shot-noise
limited signal to noise at fixed destruction.  In their simple form, KK
relations assume a linear response to the driving  fields.  Equation
(\ref{eq:limit}) includes all non-linear terms,
assuming only that the atom returns to its original state after an adiabatic
interaction with the driving fields,  a necessary condition if
we are to investigate minimally destructive processes.  In our
analysis, the non-linearities manifest themselves in the dressed state
energies.

Equation  (\ref{eq:limit}), and its extension to a cavity containing a gas
of two
level atoms, predicts that the cavity based measurement is enhanced over
the single-pass measurement by the square root of the finesse for fixed
destruction in the quantum noise limit \cite{Lye03}.   This enhacement in
signal to noise
is obtained for measurements of the transmitted or reflected  beams for
impedance matched, under-coupled and over-coupled cavities.  KK relations
between amplitude and phase (as opposed to real and imaginary components)
exist for reflection from under-coupled and impedance-matched cavities,
but do  not exist for light reflected from an over-coupled cavity, and do not
exist for transmitted light in any case \cite{KK}.  The existence of the KK relations 
for a system that does not obey our theorem shows that the limit expressed in
(\ref{eq:limit}) cannot be derived from them.  Conversely, the KK
relations cannot be generated by integrating our result over all
frequencies.

The relationship of the present work to the optical theorem also
warrants consideration. The optical theorem relates the imaginary part
of the forward scattering amplitude of a plane wave to the total
absorption cross-section. It can be applied to any
wavelike scattering from a single scattering centre, whether they be
electromagnetic waves, matter waves, acoustic waves or gravitational
waves.  The scattering event causing the decreased flux can be elastic
or inelastic.  For the scattering of electromagnetic radiation from an
extended sample of scatterers, integration of the optical theorem yields a
phase shift proportional to the real part of the forward scattering
amplitude \cite{Jackson, Lax}. This quantity cannot be related to the total
cross section. The absorption-free dispersion associated with a dark state
that motivated this work is not predicted by the optical theorem. It is
correctly described by equation
(\ref{eq:limit}) which predicts that there is no sensitivity advantage  to
such a scheme.

 \textit{Conclusions}.-
For a two level atom, the SNR for a quantum noise limited  measurement
of the column density, either via absorption or phase shift in the thin
cloud limit, depends only on destruction (spontaneous emission rate)
and bandwidth. We have shown that the use of coherent dark states or any
other combination
of coherences between levels in a multilevel atom using any number of
lasers will not improve the SNR for such a measurement. Dark states can
exhibit phase shifts that change very quickly with detuning, but any large
phase shift on either laser is always
associated with a large total excited state population and accompanying
spontaneous emission. According to the theorem derived in this Letter, any
attempt to search for a superior scheme using a more complicated level
structure will not be successful.

Although squeezed states of light or multi-pass interferometry are
experimentally challenging for (at present) moderate gains in the SNR,
they are the only ways we have found to improve on the single-pass limit
imposed by the two-level atom using classical light.  As a consequence, it
is important
that both techniques be developed.  The only alternative is to
investigate non-optical detection. Sensitive cryogenic detectors such
as SQUIDS make this an interesting possibility for any atomic species
with non-zero spin in a cryogenic environment, such as atomic hydrogen.

\begin{acknowledgments}
The Australian Centre for Quantum Atom Optics is an Australian Research
Council Centre for Excellence.  The research was supported by the
Australian Partnership for Advanced Computing.
\end{acknowledgments}

\end{document}